\documentclass[%
 preprint,
 amsmath, amssymb,
 aps, physrev, longbibliography
]{revtex4-2}

\usepackage{ hyperref}
\usepackage{rotating, graphicx}
\usepackage{color}
\usepackage{dcolumn}
\usepackage{bm}
\usepackage{xcolor}

\usepackage{bm}
\usepackage{xcolor}
\usepackage{hyperref}
\usepackage{graphicx}
\usepackage{gensymb}

\begin{document}

\title{\textbf{Data-driven  modeling of a settling sphere in a quiescent medium} 
}%

\author{Haoyu Wang$^1$\textsuperscript{\dag}, Isaac J. G. Lewis$^1$\textsuperscript{\dag}, Soohyeon Kang$^1$, Yuechao Wang$^1$, Leonardo P. Chamorro$^{1,2,3,4}$, C. Ricardo Constante-Amores$^1$}\email{crconsta@illinois.edu}
\affiliation{1. Department of Mechanical Science and Engineering, University of Illinois, Urbana, Illinois 61801, USA.\\
2. Department of Aerospace Engineering, University of Illinois, Urbana, Illinois 61801, USA.\\
3. Department of Civil and Environmental Engineering, University of Illinois, Urbana, Illinois 61801, USA.\\
4. Department of Earth Science and Environmental Change, University of Illinois, Urbana, Illinois 61801, USA.
}
\thanks{\textsuperscript{\dag}These authors contributed equally to this work.}

\begin{abstract}
We develop data-driven models to predict the dynamics of a freely settling sphere in a quiescent Newtonian fluid using experimentally obtained trajectories. Particle tracking velocimetry was used to obtain a comprehensive dataset of settling motions, which we use to train neural networks that model the spatial evolution of a spherical particle without explicitly resolving the surrounding fluid dynamics. We employ deterministic neural ordinary differential equations (NODEs) and stochastic neural stochastic differential equations (NSDEs) to reconstruct the sphere’s trajectory and capture key statistical features of the settling process. The models are evaluated based on short- and long-time dynamics, including ensemble-averaged velocity evolution, settling time distributions, and probability density functions of the final settling positions. We also examine the correlation between lateral displacement and streamwise velocity and assess the impact of dataset size on predictive accuracy. While NODEs excel in trajectory reconstruction and generalization across different initial conditions, NSDEs effectively capture statistical trends in the long-time behavior but are more sensitive to data availability. Acceleration profiles computed via second-order finite difference schemes confirm that both approaches accurately capture long-time dynamics, though short-time transients pose challenges.
\end{abstract}

\maketitle


\section{Introduction}

\begin{figure}
    \centering
    \includegraphics[width=0.99\linewidth]{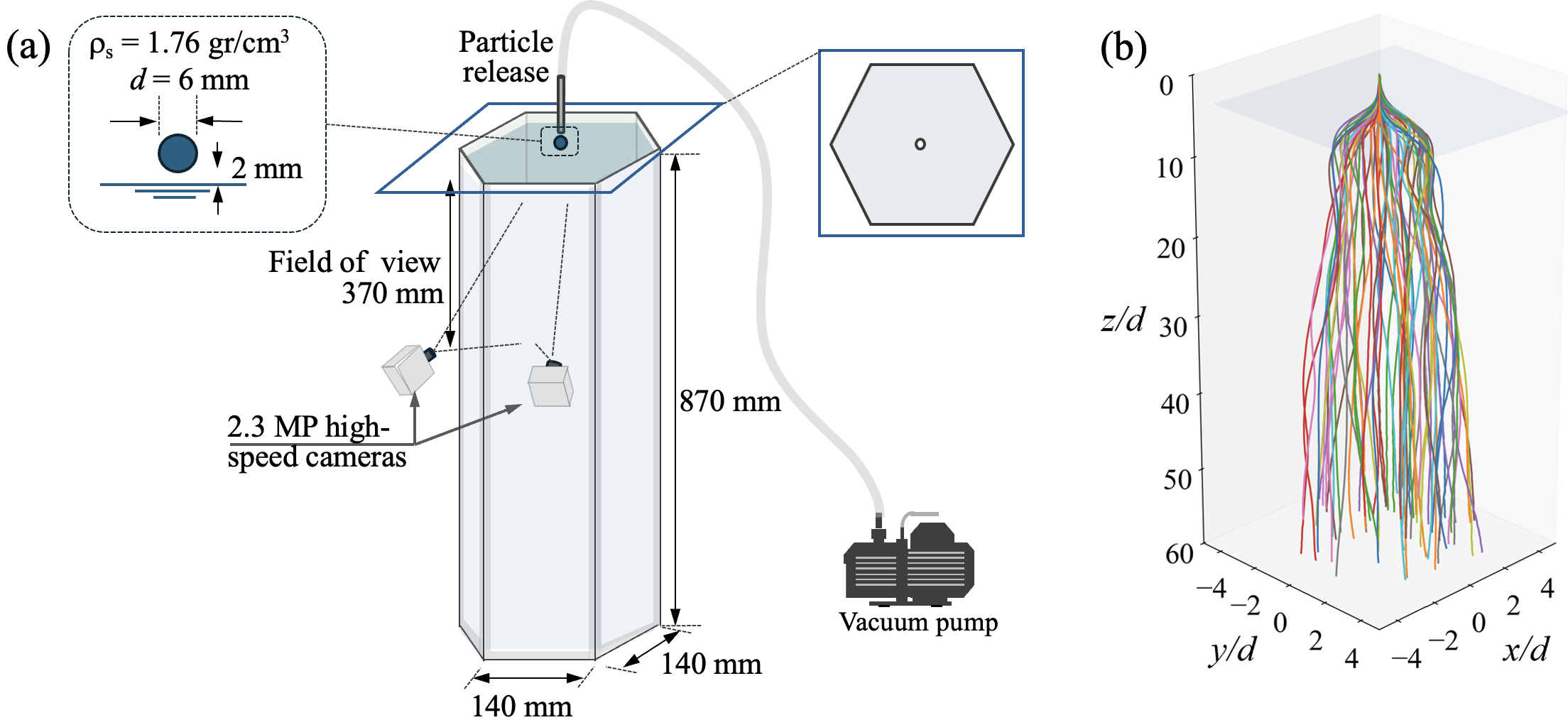}
    \caption{\label{fig:setup}(a) Basic schematic of the experimental setup, (b) representative particle trajectories using PTV.}
\end{figure}
The settling dynamics of a spherical particle in a quiescent medium is a canonical problem in fluid mechanics ubiquitous in environmental and industrial systems across a range of scales. In the environment, the settling of particulate matter affects the morphological state and evolution of riverbeds, estuarine networks, and coastal environments by modulating bedload and suspended load fluxes \cite{dubuis2023clogging, winterwerp2002flocculation, oursel2014behaviour}, while the settling of biological particles is critical for ecosystem equilibrium \cite{maggi2013settling, yoon2022biodegradable, marcharla2024microplastics}. The interplay between various factors, including inertial, viscous, and buoyancy forces and the characteristics of the medium and flow field, governs the Lagrangian dynamics of particle depositional patterns. In pollutant dispersion, the settling of aerosols, microplastics, and active particulates is a particularly critical problem. Industrial applications, such as wastewater treatment, pharmaceutical processes, and developing new-generation sensing technologies \cite{saravanan2021effective, ulusoy2023review, goral2023settling, shetty2024experimental, bousiotis2023monitoring}, demand robust statistical approaches to determine particle settling. The underlying physics often involves wake dynamics, vortex shedding, and turbulence-particle interactions, requiring a combination of theoretical, numerical, and experimental methodologies to develop predictive models capable of capturing the statistical and mechanistic aspects of settling behavior under general flow conditions.

The settling behavior of spherical particles has been extensively studied both experimentally and numerically \citep{Allen01111900,KARAMANEV01051996, Truscott_Epps_Techet_2012}. A particle under that state can exhibit different trajectory regimes once the axisymmetric wake becomes unstable. \citet{jenny2004} showed that the dynamics of path instability depend on two key parameters, the solid-to-fluid density ratio $\rho_s/\rho_f$, and Galileo number $Ga = \nu^{-1} (\lvert \rho_s/\rho_f-1 \rvert gd^3)^{1/2}$, representing the ratio between the gravity-buoyancy and viscosity effects; here, $\rho_s$ and $\rho_f$ are the particle and fluid densities, $g$ denotes gravitational acceleration, $d$ is the particle diameter, and $\nu$ is the kinematic viscosity of the fluid. Their numerical simulations for $150 < Ga < 350$ and $0<\rho_s/\rho_f<10$ showed that a free sphere followed a straight trajectory when $Ga < 156$ with an axisymmetric wake. As the $Ga$ increased, a steady oblique trajectory appeared with plane-symmetric wake structure, which was followed by an oscillating oblique regime at higher $Ga$. For comparatively heavier spheres, a further increase in $Ga$ led to chaotic regime with significant three-dimensional trajectory variability. In the experimental study by \citet{veldhuis2007}, the characterization of particle trajectories with visualizations of the wake structure using the Shlieren technique showed a general agreement with the results of \citet{jenny2004}. \citet{horowitz2010} performed an extensive experimental study of settling spheres for $100 < Ga < 20000$ and $0 < \rho_s/\rho_f < 1.5$. They examined wake structures  in each regime using fluorescent dye visualization along with particle image velocimetry (PIV). For $\rho_s/\rho_f > 0.6$, periodic vortex shedding was observed in the oscillating oblique regime. When $Ga > 380$, the oblique motion became intermittent with vortex rings shedding in varying orientations. For a rising sphere, they pointed out that its trajectory is highly sensitive to background disturbances, appearing rectilinear in quiescent fluid but exhibiting relative large random transverse motions under small fluid disturbances. \citet{zhou2015} extended the numerical study by \citet{jenny2004} covering the the higher $Ga$ in the chaotic regime ($250 < Ga < 500$). They observed that for $\rho_s/\rho_f > 1.7$, the transition to chaotic state occurred subcritically, with coexisting states, one maintaining a fixed symmetry plane and another with the plane slowly rotating. After this bistable regime, three-dimensional trajectories become fully chaotic, often forming nearly helical paths. Given the existence of random natural perturbation and the nonlinear interaction between the settling object and the fluid medium, it is generally challenging to obtain an analytical formulation to fully describe the characteristics of the object settling system.

Data-driven modeling may offer predictive solutions for a range of applications. It has been explored to study related problems, including the falling of multiple flexible filaments \citep{fox_sedimenting}, sediment settling velocities in fluids \citep{DELCEY2024117389,2019AGUFMEP51F2184C}, and in estimating clumping and settling dynamics using particle image velocimetry data in water treatment applications \citep{BRESSANE2024106138}. 
These studies primarily focused on learning the long-time dynamics, such as settling time, terminal velocity, and drag coefficient \cite{song2017new, xu2019settling, qian2024machine, lv2024prediction, keren2024improved}.
While long-time dynamics can provide valuable insights, they do not capture the full range of dynamical behavior observed in experiments, particularly where short-time interactions strongly impact the long-time outcome.

When the governing equations are unknown and only temporal snapshots of data are available, the process can be assessed as a dynamical system. If a coordinate system is represented by \({\bf{x}} \in \mathbb{R}^{d_N}\), the corresponding evolution equation in these coordinates is given by \(d{\bf{x}} /dt=g({\bf{x}})\).  For low-dimensional systems, when time derivative data is available, the sparse identification of nonlinear dynamics (SINDy) method is commonly used \cite{Brunton}. SINDy applies sparse regression on a dictionary of candidate terms to discover the governing vector field and is particularly effective for systems with relatively simple dynamics. A more general approach, the neural ordinary differential equation (NODE) framework \cite{chen2019neural}, represents $g$ as a neural network and does not require derivative data. NODE has been successfully applied to complex chaotic systems \cite{alec_couette,alec_chaos,edmd_dl_ad,pipe_flow,disdmand},  and will be employed in this work. Finally, other state-of-the-art data-driven frameworks for forecasting can be achieved using reservoir networks \cite{Pathak, Doan, Fan, Platt, Racca}, recurrent neural networks \cite{Vlachas}, transformers \cite{Geneva, Gilpin}, and quantum reservoir computing \cite{Ahmed2024}. These approaches rely on discrete-time maps, are inherently non-Markovian, and often expand the dimensionality of the state space to capture a richer feature representation of the system's true dynamics. 

In this study, we develop a data-driven approach to predict the dynamics of a settling sphere (interchangeable as particle) in a quiescent Newtonian medium. Rather than modeling the underlying fluid dynamics, we directly learn the evolution of the particle's spatial location from experimental data. To our knowledge, such datasets are rare as they involve high-resolution tracking a sphere released from rest right above a free surface, tracking its motion over the distance required to achieve terminal velocity. We will train neural networks to learn the sphere's evolution using both deterministic and stochastic approaches. We demonstrate that these models can accurately forecast the time evolution with minor error. We compare between the benefits of two types of neural ODE architectures: the classical neural ODE and the stochastic neural ODE.  

In Section 2, we describe the experimental data collection process and the framework for modeling time evolution. In Section 3, we present the results, including model predictions for both short-time tracking and long-time statistics. Finally, in Section 4, we summarize our key findings and conclusions.


\section{Methodology \label{method}}
\subsection{Experimental setup \label{exp_data}}

Single spherical particles were released individually at the center and close to the free surface of a hexagonal tank filled with quiescent water maintained at 25$^\circ$C. The tank, constructed with transparent walls for optical access, had a height of 370 mm and a side width of 140 mm. These dimensions ensured minimal wall effects, allowing the observed particle dynamics to closely approximate unbounded conditions.

The particles had a diameter of $d=6$ mm and a density of $\rho_s=1.76$ g cm$^{-3}$, resulting in a density ratio of $\rho_s/\rho_w = 1.77$ and a Galileo number of $Ga = 1428$, where $\rho_w\approx 997$ g cm$^{-3}$. Under these conditions, particles exhibit three-dimensional chaotic trajectories during their settling. To ensure controlled and repeatable initial conditions, each particle was positioned approximately 2 mm above the free surface using a vacuum-based release mechanism. Upon deactivation of the vacuum chamber, the particle was released from rest, ensuring negligible initial velocity. This method effectively minimized uncontrolled perturbations at release, allowing robust repeatability. The experiment was repeated approximately 200 times, with each trial conducted after allowing the fluid to return to quiescence. All trajectories originated from a common initial position, defined as $(x/d, y/d, z/d)=(0,0,0)$. A schematic of the experimental setup is illustrated in Fig. \ref{fig:setup}a.

The three-dimensional trajectories of the particles were tracked using a customized dual-camera Particle Tracking Velocimetry (PTV) system \citep{kang2024}. Two high-speed 2.3 MP cameras, operating at a sampling frequency of 150 Hz, were positioned to capture two adjacent faces of the hexagonal tank. At each time step, the center positions of the particles were extracted from image sequences, and their three-dimensional trajectories were reconstructed using a triangulation procedure. This reconstruction was performed by computing the intersection of rays originating from both cameras, with ray directions determined by the respective camera calibration matrices. These matrices, incorporating intrinsic parameters such as focal length and lens distortion coefficients, were obtained via a checkerboard calibration procedure following the method of \citet{zhang2000}. The accuracy of the 3D reconstruction was assessed using the re-projection error, which quantifies the discrepancy between the original and re-projected pixel coordinates of the tracked particles. The average re-projection error corresponded to $0.07d$. Fig. \ref{fig:setup}b shows a sample of representative superimposed particle trajectories.

\subsection{Data-driven modeling}

\subsubsection{NODE: Deterministic modeling \label{node}}
Let's assume a deterministic, Markovian dynamics, and 
$\boldsymbol{x}\in \mathbb{R}^{d_N} $ represent the state, which is  defined as the position of the particle center  in the $(x,y,z)$ coordinate system. 
A common approach in data-driven modeling is to approximate the time evolution of $\bm{x}$ using a learned function $\boldsymbol{g}$  as follows
\begin{equation}
\frac{d\bm{x}}{dt}=\bm{g}(\bm{x}).
\label{eq:NODE}
\end{equation}
here,
$\boldsymbol{g}$  is the vector field, and we use a NODE framework. Representing $\boldsymbol{g}$ as a neural network with weights $\theta_f$, we can time-integrate Eq.~\ref{eq:NODE}  between $t$ and $t+\delta t$ to yield a prediction $\tilde{\boldsymbol{x}}(t+\delta t)$, such as
\begin{equation}
\tilde{{\boldsymbol{x}}}(t+\delta t)=\boldsymbol{x}(t)+\int_{t}^{t+\delta t}\boldsymbol{g}(\boldsymbol{x}(t');\theta_f) dt'.   \label{eq:NODE2}
\end{equation}
Given data for $\boldsymbol{x}(t)$ and $\boldsymbol{x}(t +\delta t)$ for a long time series we can train $\boldsymbol{g}$  to minimize the $L_2$ difference between the prediction $\tilde{\boldsymbol{x}}(t+\delta t)$  and the known data  $\boldsymbol{x}(t+\delta t)$. Then the mean squared error (MSE) loss function is defined by
\begin{equation}
\mathcal{L}_g=\frac{1}{B}\sum_{i=1}^B||\boldsymbol{x}(t_i+\delta t)-\tilde{\boldsymbol{x}}(t_i+\delta t)||_2^2.
\label{loss_node}
\end{equation}
where \(B\) is the batch size. We use automatic differentiation to determine the derivatives of $\boldsymbol{g}$ with respect to  $\theta_f$ \citet{chen2019neural}.  To optimize the loss function described in equation \ref{loss_node}, we use an AdamW optimizer in PyTorch \citet{pythorch}. Table 1 
summarizes the network architectures used in this study for the NODE.

Finally, while stabilization terms have been found beneficial in some applications of NODEs for complex dynamical systems \citet{alec_chaos}, we did not find them necessary in this study.

\subsubsection{NSDE: Stochastic modeling \label{nsde}}

\begin{table}
    \centering
    \caption{\label{table_arch}Neural network architectures for the  NODE and NSDE. `Shape' represents the dimension of each layer, `Activation'  refers to the types of activation functions used. }
    \begin{tabular}{ccccc}
        \hline
        Approach& Function & Shape & Activation  & Learning Rate \\
        \hline
        NODE &  $\boldsymbol{g}$	    &  $d_N/64/$$d_N$ \quad  & GELU/lin & $[10^{-3},10^{-4}]$ \\
        NSDE & $\boldsymbol{g}$	    &  $d_N$/64/$d_N$ \quad  & LipSwish/lin & $[10^{-3},10^{-4}]$ \\
        \hline
    \end{tabular}
\end{table}

Due to the nature of the experimental data gathered and the underlying physical processes, a promising modification to  classical deterministic data-driven models is to represent the system by a stochastic differential equation (e.g.,   due to potential noise in experimental setups).
In particular,  the time evolution of the data can be represented as
\begin{equation}
dX_t=f(t,X_t)dt+g(t,X_t)dW_t,
\end{equation}

where $X_t$ is a stochastic process to be modeled, $f$ is a function describing the deterministic part of the differential equation, commonly referred to as  the drift function, and $g$ is a diffusion function describing the variance of the solution with respect to a Wiener process $W_t.$ When $g$ is set to zero, the SDE collapses to an ODE, as the absence of a diffusion term implies a purely deterministic evolution. 
In contrast, the presence of the stochastic process 
$W_t$ introduces randomness into the dynamics, allowing the model to account for both measurement uncertainty and sensitivity to initial conditions.
$W_t$ can be described  as a Brownian motion $B_t,$ which satisfies the properties required required for both  Ito and Stratonovich integrals. Then, the solution to the SDE can be expressed as 
\begin{equation}
X_t = X_0+\int_0^tb(\tau,X_{\tau})d\tau+\int_0^t\sigma(\tau,X_{\tau})\circ dB_{\tau},\label{eq:sde_solution}
\end{equation}
where $X_0$ is an initial distribution for the random variable $X_t$, and the operator $\circ$ on the differential $dB_\tau$ implies Stratonovich integration. The difference between Ito and Stratonovich integration lies only in the difference of time evaluation of the Brownian motion term. In particular, considering a function $f(t,\omega),$ where $\omega\in\Omega\subset\mathbb{R}^n$ is an element of the dataset, the integral 
\begin{equation}
\int_{t_0}^Tf(t,\omega)dB_t
\end{equation}
is evaluated as the limit as $n\rightarrow\infty$ in the interval $[t_0,T]$ of
\begin{equation}
\sum_{j=0}^{n}f(t^*_j,\omega)(B_{t_{j+1}}-B_{t_j}).\label{eq:integ_lim}
\end{equation}
\\
Note that Eq.~\ref{eq:integ_lim} resembles a Riemann–Stieltjes integral. Here, the notation $\circ$ demarcates the use of Stratonovich integration, which commonly works well with numerical evaluation of stochastic integrals. The only difference in evaluation of the Ito and Stratonovich integrals is in the choice of $t^*_j.$ In Ito integration, $t^*_j$ is chosen as $t_j,$ the left point of each subinterval, and in Stratonovich integration, $t^*_j$ is chosen to be the midpoint, $t^*_j=(t_{j+1}-t_j)/2.$
\\
To model the SDE using a neural differential equation requires the neural network to learn the time evolution with respect to the noisy behavior of the solution (see  \citet{kidger2022}). To do this, we model the NDE with two separate networks that represent the drift and diffusion terms. In this formulation, the model becomes
\begin{equation}
dX_t = f_{\theta}dt+g_{\theta}dB_t,
\end{equation}
\\
here $f_{\theta}$ and $g_{\theta}$ are   neural networks describing the drift and diffusion terms, respectively, where $\theta$  denotes the set of learnable network parameters. The term  $dB_t$ corresponds to artificially generated Brownian noise. The neural SDE is implemented using two separate  multi-layer perceptrons, one which is reserved for learning $f$ and the other which is reserved for learning.
The approximate solution to the SDE in the range $t\in[t_0,t_1]$ is then described as
\begin{equation}
X_{t_1} = X_{t_0}+\int_{t_0}^{t_1}f_{\theta}(X_t,t;\theta)dt+\int_{t_0}^{t_1}g_{\theta}(x_t,t;\theta)dB_t.
\end{equation}
Determination of the approximate solution with respect to the neural network must be done using an appropriate SDE solver. There are multiple solvers which can be used in backpropagation of the neural network which have varying degrees of computational efficiency and accuracy, including the Euler-Murayama, Milstein, Reversible Heun, and Euler-Heun. In our analysis, SDE solutions are determined using the Euler-Murayama scheme using the torchsde python package.

Due to the fact that the solution of the SDE is inherently a stochastic process $X_t,$ individual trajectories, as opposed to pure probabilistic behavior, poses a significant challenge in the NSDE framework. To address this, a control mechanism is often required to relax  the stochastic behavior onto  a more well-defined path. 
One approach  is  to model the neural network using a generative adversarial network (GAN), first defined in \citet{goodfellow2014generativeadversarialnetworks}. GANs consist  of two neural networks, termed the generator and discriminator. Unlike standard feedforward neural networks which employ backpropagation through the hidden state in order to minimize a loss function, such as the mean-squared error, the generator provides solutions from a distribution whose statistics match that of the incoming data, while the discriminator attempts to determine how likely that solution is to be from the dataset. Hence, the loss to be minimized is instead the Wasserstein norm between the predicted probability space and the true probability space.
In the GAN framework, the generator is the network approximating the SDE, while the discriminator is defined as a neural controlled differential equation (NCDE), which serves to discriminate trajectories belonging to the true dataset from the set of those generated by the NSDE \cite{kidger2021}. However, it is not entirely necessary to model the NSDE as a GAN. In our work, we adopt the classical neural SDE from \citet{oh2025stableneuralstochasticdifferential} which directly employs the control term inside of a single neural network defining the approximate equation of motion of the SDE. Specifically, this control term is introduced as
\begin{equation}
\bar{X}_t=\xi(t,X_t,\chi(t);\theta_{\xi}),
\end{equation}
where $\chi(t)$ is a controlled path and $\xi$ is a neural network parameterized by weights $\theta_{\xi}$. By defining such a function, we are able to better learn the evolution of trajectories given by the input data through the drift term $f_{\theta}$. As shown in \citet{oh2025stableneuralstochasticdifferential}, this addition of a control term has been shown to guarantee the existence of a unique strong solution, which is necessary in increasing stability of the SDE solver, and subsequently the robustness of training.  This control-based formulation offers several advantages over the GAN formulation, in particular, it avoids the need for a separate discriminator network, enabling to work within a single latent space.
\\
The definition of the controlled path $\chi(t)$ is done through the use of some interpolation scheme. Two common methods used in NSDEs are linear interpolation schemes and hermite cubic splines, the latter of which is defined in \citet{morrill2021neuralcontrolleddifferentialequations} with the consideration of time-series data which has been irregularly sampled. By defining our network $\xi$ with respect to the cubic spline coefficients, the continuously defined paths only need to be pre-initialized while still giving the advantage of a unique strong solution throughout training of the SDE.
\\
The use of such a control term as the cubic spline path is additionally motivated by theory from neural controlled differential equations (NCDEs) for which the evolution equation is defined as
\begin{equation}
    dz(t)=\phi(t,z(t))dX(t),
\end{equation}
where $X(t)$ is some driving function which need not be a stochastic process. Such a differential equation model is generally applicable when the time-dependent behavior of some physical process changes due to additional time-dependent forcing. Many examples of such situations are in medical applications \citep{NEURIPS2020_4a5876b4}, where the likelihood of the onset of certain pathologies and their effects depend on certain time-dependent variables such as age.
\\
The solution of a CDE is expressed as
\begin{equation}
z(t_1)=z(t_0)+\int_{t_0}^{t_1}\phi(\tau,z(\tau))dX_{\tau},
\end{equation}
and hence the definition of the solution with respect to the NCDE is
\begin{equation}
z(t_1)=z(t_0)+\int_{t_0}^{t_1}\phi_{\theta_c}(\tau,z(\tau))dX(\tau),
\end{equation}
given a neural network $\phi_{\theta_c}$ parameterized over CDE network weights $\theta_c.$ In the setting of continuous-time representations of neural networks which are required of neural differential equations, we would like the term $X(t)$ to also be defined continuously over time. The most straightforward way to do this is by recasting $X(t)$ as an interpolated path. 
\\
In the case of NSDEs,  \citet{oh2025stableneuralstochasticdifferential} demonstrated that incorporating a  control term yields better performance than models that omit such a term.
In particular,  the enforcement of smoothness through the control term
enables  a more effective handling of time-dependent stochasticity in the system dynamics. While stochasticity  arising from measurements  is  time-independent, the uncertainty intrinsic to dynamics themselves may vary over time and is not guaranteed to follow such assumptions. 
The inclusion of a control term provides  an extra degree of freedom that allows the model to account  for this non-stationary behavior, which would be difficult to capture with standard SDE.
As a result, the model can  balance the stochasticity of the system with the smoothness of the control term, enhancing robustness   to the rougher trajectories often introduced by   experimental data and to variability arising from complex underlying dynamics. Despite the addition of stochasticity, however, the loss to be minimized in this particular SDE framework is the mean-squared error. In particular, the optimization problem is to minimize the mean-squared error of the reconstructed trajectories:
\begin{equation}
L(X_t)=\left\lVert X_{t,\text{true}}-X_{t,\text{pred}}\right\rVert_2^2,
\end{equation}
where $X_{t,\text{true}}$ and $X_{t,\text{pred}}$ are the true and predicted stochastic processes, respectively, and $\lVert\cdot\rVert_2$ is the $L^2$ norm. Additionally, the predicted solution is given by the integral in \ref{eq:sde_solution} using the appropriate numerical SDE solver. We do not aim for the losses to approach zero, as we would like the model to generate new trajectories whose statistics closely match that of the true data without encountering overfit.
\\
We remark further that more advanced neural SDE frameworks have been developed and benchmarked in \citet{oh2025stableneuralstochasticdifferential}. These networks include neural network models of the linear multiplicative noise SDE (LNSDE), geometric noise SDE (GSDE), and Langevin-type SDE (LSDE). As the names suggest, the neural LNSDE considers the diffusion term to be
\begin{equation}
\sigma(t;\theta_{\sigma})z(t)dB_t,
\end{equation}
while the neural GSDE considers the diffusion term as
\begin{equation}
\sigma(t;\theta_{\sigma})dB_t,
\end{equation}
which does not include the latent variable $z(t)$. The LSDE considers the drift term to be autonomous, i.e., one of the form
\begin{equation}
f_{\theta}(X_t;\theta_{f}).
\end{equation}
We do not consider these frameworks in this paper, as there is no immediate justification to treat the drift term as autonomous in the present physical system, and the noise which arises due to measurement uncertainty should not be treated as multiplicative or geometric. Also, the treatment of noise as diagonal reduces the number of degrees of freedom necessary for numerical solution of the SDE, hence greatly increasing computational efficiency.

In training, we use the Lipswish activation function proposed by \citep{chen2020}:
\begin{equation}
    \text{LipSwish}(x)=\frac{x\sigma(\beta x)}{1.1}.
\end{equation}
\allowbreak
Here, $\sigma$ is the sigmoid function, $\sigma(x) = 1/(1+e^{-x})$, and $\beta$ is a learned parameter consistent with the definition of the Swish function, $\text{Swish}(x)=x\sigma(\beta x)$. During training, $\beta$ is set to one, and the term $x\sigma(x)$ is implemented using the SiLU function from PyTorch. The constant $1/1.1$ is used to ensure the Lipschitz condition, i.e., a first derivative bounded by 1 for all $x$in the domain \citep{chen2020}. Enforcing the Lipschitz condition in a neural network enhances its  robustness to data perturbations  \citep{scaman2019}. 
Automatic differentiation was used to backpropagate through the data, and an Adam optimizer  was used to minimize the loss function over the weight parameters. Additionally, PyTorch was used to create the NSDE model. 
The neural network structure was adapted from the Neural SDE framework given in \citep{oh2025stableneuralstochasticdifferential}. Table 1 summaries the network architectures used in this study for the NSDE.

\section{Results and discussion}
\label{results}
This section presents the NODE and NSDE approaches on the experimental data described in Section \ref{exp_data}. We show how these methods are capable of predicting the collective behavior of the system, and evaluate their performance in both short-time and long-time statistics. 

During the course of optimizing the neural networks, multiple hyperparameters including the batch size, hidden dimension, number of network layers, and the learning rate were systematically varied. We observed that keeping the  network small was essential, as increasing the number of layers and the hidden dimension led to overfit. In the NSDE, we also evaluated the impact of dropout in the hidden layers, finding that it substantially increased computational cost while providing negligible improvement in generalization or training performance. Both models described in Table~\ref{table_arch} were trained on a single compute node with 8 cores of an AMD EPYC 7763 “Milan” processor with 16 GB (for NODE) or 64 GB (for NSDE) of memory. The NODE model was trained over 48,000 iterations with a batch size of 256, which takes about 1 hour to finish; the NSDE model was trained over 30,000 iterations with a batch size of 16, requiring approximately 12 hours to finish. For NODE, all the experimental trajectories were first randomly split into 80$\%$ and 20$\%$ as the training and test sets, respectively. Each trajectory was then segmented into short sequences, where each sequence contained a fixed number of consecutive, experimentally sampled trajectory points, controlled by a hyperparameter 'batch time'. We additionally note that the batch size of NODE means the number of short trajectory sequences per batch rather than full trajectories. This contrasts with the batch size used in NSDE, which corresponds to the number of full trajectories sampled per batch. Each trajectory is represented by its spline coefficients, which are passed to the neural network for learning. As such, the NSDE batch size of 16 which was used in reconstruction in Section \ref{results} selects approximately 8.5\% of the data for each mini-batch to be used in the stochastic gradient descent algorithm. This batch size provided a good balance between computational efficiency and model performance. Additionally, the NSDE model was trained on the full dataset for the 100\% data case, while for the 50\% and 20\% cases given in \ref{results}, training and test sets were split 50\% for training, 50\% for testing, and 20\% for training, 80\% for testing, respectively. Besides, in both architectures, we adjusted the learning rate dynamically during training by using a StepLR scheduler to ensure stable convergence. We started with a value of \(10^{-3}\) and a weight decay factor of \(10^{-3}\) in NODE. The NSDE did not use weight decay. After two-thirds of the total training iterations, the learning rate was reduced by an order of magnitude, reaching the final learning rate of \(10^{-4}\).


\subsection{Reconstruction by NODE \& NSDE}

First, we assess and compare the accuracy of the NODE-based and NSDE-based predictions as described in Section \ref{node} and \ref{nsde} by comparing the individual particle trajectories to the experimental data.  For the NODE case, the initial conditions for the reconstructed trajectories are sampled from a fitted normal distribution, with the mean and variance determined from the initial conditions of the full experimental dataset. A total of 400 initial conditions are randomly sampled for trajectory reconstruction. By sampling new initial conditions rather than using the actual test set, we are performing an ensemble-based validation, focusing on comparing the statistical properties of the predicted trajectories to the ground truth, rather than direct one-to-one comparisons. This approach demonstrates the model's ability to capture the collective behavior of the system and generalize on unseen initial conditions. In Appendix A1, we present the reconstruction of the trajectories from the NODE model using initial conditions derived solely from the test data. However, because the control term in the NSDE architecture requires prior generation of cubic spline coefficients, it was necessary to use only experimentally gathered data for the NSDE case. As a result, analysis done by the NSDEs consisted of exactly 189 trajectories as opposed to the 400 randomly generated initial conditions gathered for the NODE case described above. 

\begin{figure}
    \centering
    \includegraphics[width=1\linewidth]{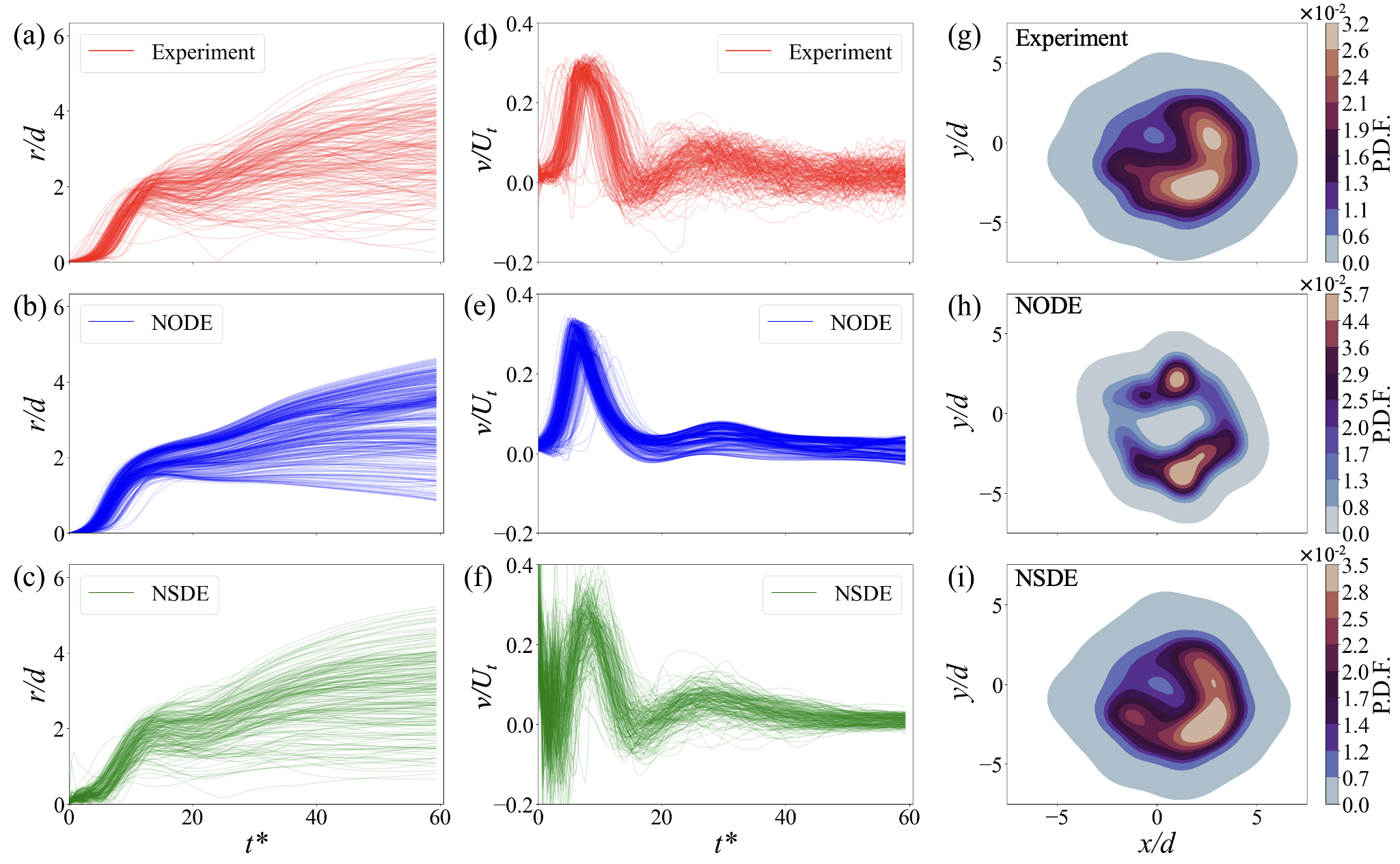}
    \caption{\label{lateral_pdf} (a-c) Dimensionless lateral displacement \(r/d\) of each trajectory as a function of the dimensionless time \(t^*\). (d-f) Dimensionless lateral velocity $v/U_t$ ($U_t=363$ mm/s is the terminal velocity) of each trajectory as a function of $t^*$. (g-i) Probability density maps of horizontal settling positions in the horizontal $(x-y)$ plane at $t^* \sim 56$.}
\end{figure}

Fig. \ref{lateral_pdf}a-f illustrates the temporal evolution of the lateral displacements and velocities for both the experimental and predicted trajectories. There, the dimensionless time is $t^*=tU_t/d$ and $U_t$ is the terminal settling velocity. We combine the horizontal coordinate pair $(x,y)$ into a unified lateral coordinate $r$ representing the horizontal Euclidean distance from the origin, i.e., $r=\sqrt{x^2+y^2}$. Then, the streamwise and lateral velocity pair $(u, v)$ is computed using the finite difference scheme with a time step of $\delta t$:
\begin{equation}
   \begin{bmatrix}
u \\
v \\
\end{bmatrix} =\mathbf{v}(t) \approx \frac{\mathbf{x}(t +\delta t)-\mathbf{x}(t-\delta t)}{2\delta t},\mathbf{x}=\begin{bmatrix}
z \\
r \\
\end{bmatrix}
\end{equation}
where $z$ is the streamwise coordinate representing the settling depth from the origin.

\begin{figure}
    \centering
    \includegraphics[width=1\linewidth]{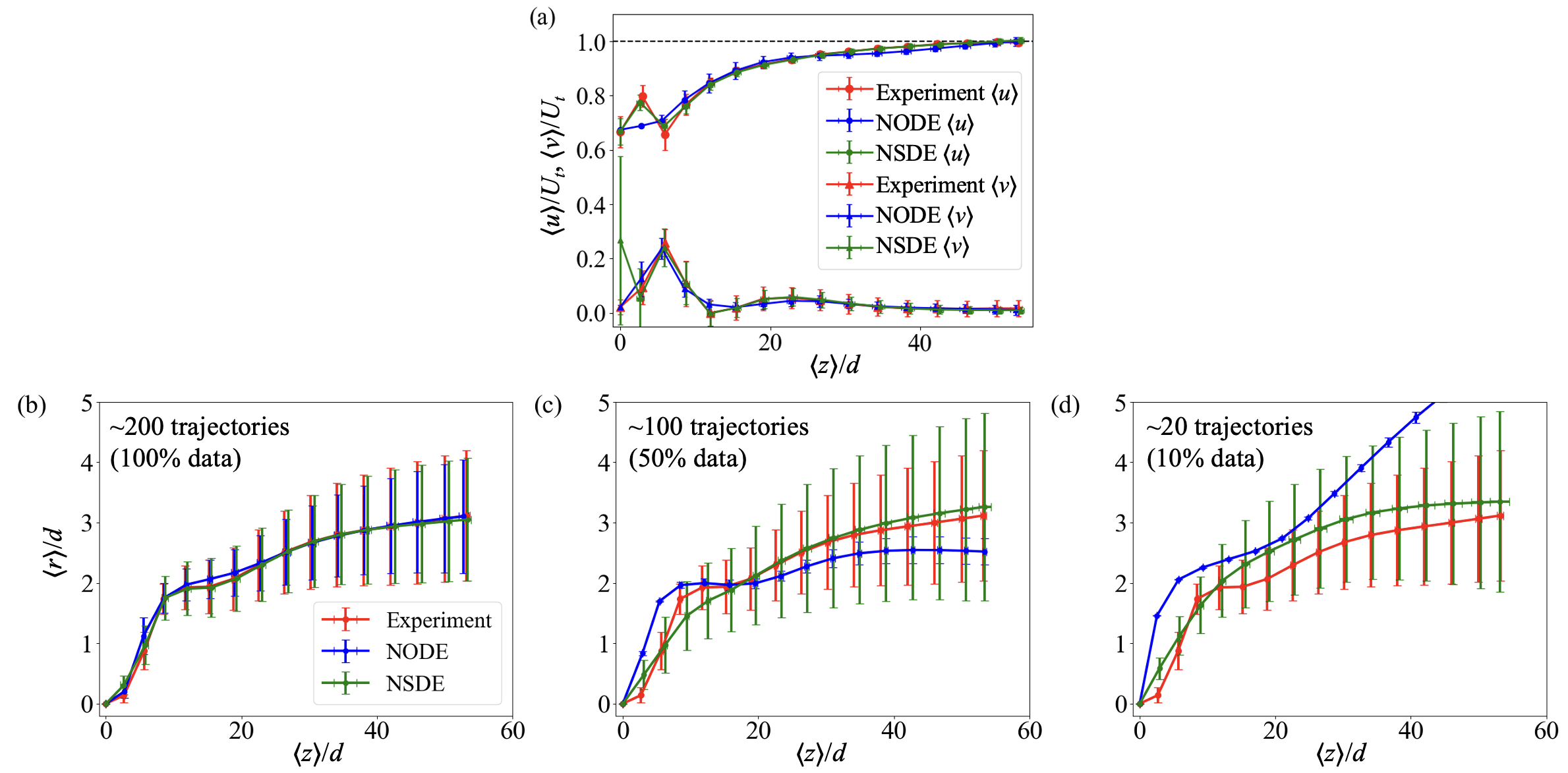}
    \caption{\label{means}(a) Experimental and reconstructed ensemble-averaged streamwise, $\langle u \rangle$, and lateral, $\langle v \rangle$, velocities as functions of dimensionless settling depth $\langle z \rangle/d$, normalized by terminal velocity $U_t$. Vertical and horizontal error bars indicate velocity and depth standard deviations. (b–d) Experimental and reconstructed ensemble-averaged dimensionless lateral displacement $\langle r \rangle/d$ as a function of $\langle z \rangle/d$. NODE models are trained with $100\%$, $50\%$, and $10\%$ of the dataset. Error bars represent standard deviations of displacements and depths.}  
\end{figure}

Initially, the stationary sphere is released into the air and  accelerates due to gravity leading to an increase in streamwise velocity. As the sphere penetrates the air-fluid interface, turbulent wake effects result in lateral motion in an arbitrary direction (see Fig. \ref{lateral_pdf}a \& d). Once fully submerged, buoyant and drag forces begin to decelerate the sphere in both streamwise and lateral directions. Around $t^*\sim 8$, the lateral velocity reaches its peak before decaying to zero at approximately the same rate as in the acceleration phase. Fig. \ref{lateral_pdf}b \& e demonstrate that the NODE model accurately capture the temporal evolution of lateral displacements and velocities, reproducing both the settling behavior and the rate at which lateral motion decays in individual trajectories; however, in Fig. \ref{lateral_pdf}c \& f, the NSDE predicted trajectories fail to capture the initial transient stage and the lateral velocity peak becomes less identifiable. Finally, Fig. \ref{lateral_pdf}g-i show probability density maps in the horizontal $(x-y)$ plane at $t^*\sim56$, indicating that the NSDE model outperforms the NODE model in statistically recovering particle locations at the long-time terminal stage. While some discrepancies in the NODE predictions are observed, these can be attributed to the inherent complexity and sensitivity of the particle's motion to initial conditions, as well as its initial interaction with the free surface, even in a quiescent medium. 

Next, we  evaluate the performance of the models regarding the reconstruction of the collective statistical characteristics of the  system  as a function of depth, using the trajectories from Fig.  \ref{lateral_pdf}. 
The collective properties are obtained by calculating the ensemble-averaged values of the streamwise and lateral velocities, as well as the lateral displacement, at each selected depth, using
\begin{equation}
    \langle u \rangle(z)=\frac{1}{N}\sum_{i=1}^N u_i(z),
\end{equation}
\begin{equation}
    \langle v \rangle(z)=\frac{1}{N}\sum_{i=1}^N v_i(z),
\end{equation}
\begin{equation}
    \langle r \rangle(z)=\frac{1}{N}\sum_{i=1}^N r_i(z),
\end{equation}
where $\langle u \rangle$ and $\langle v \rangle$ are the ensemble-averaged streamwise and lateral velocity; $\langle r \rangle$ is the ensemble-averaged lateral displacement, $i$ denotes the index of the corresponding settling trajectory, and $N$ is the total number of trajectories taken into consideration. As presented in Section \ref{exp_data}, the discrete sampling introduces challenges in extracting position data at the precisely  selected depths. To address this, and assuming a strong temporal correlation between the streamwise displacement and time evolution in this system, we adopt an alternative approach. Specifically, we select a series of evenly spaced time steps and compute the ensemble-averaged depths of trajectories at these instances, using
\begin{equation}
    \langle z \rangle(t)=\frac{1}{N}\sum_{i=1}^N z_i(t).
\end{equation}


\begin{figure}
    \centering
    \includegraphics[width=1\linewidth]{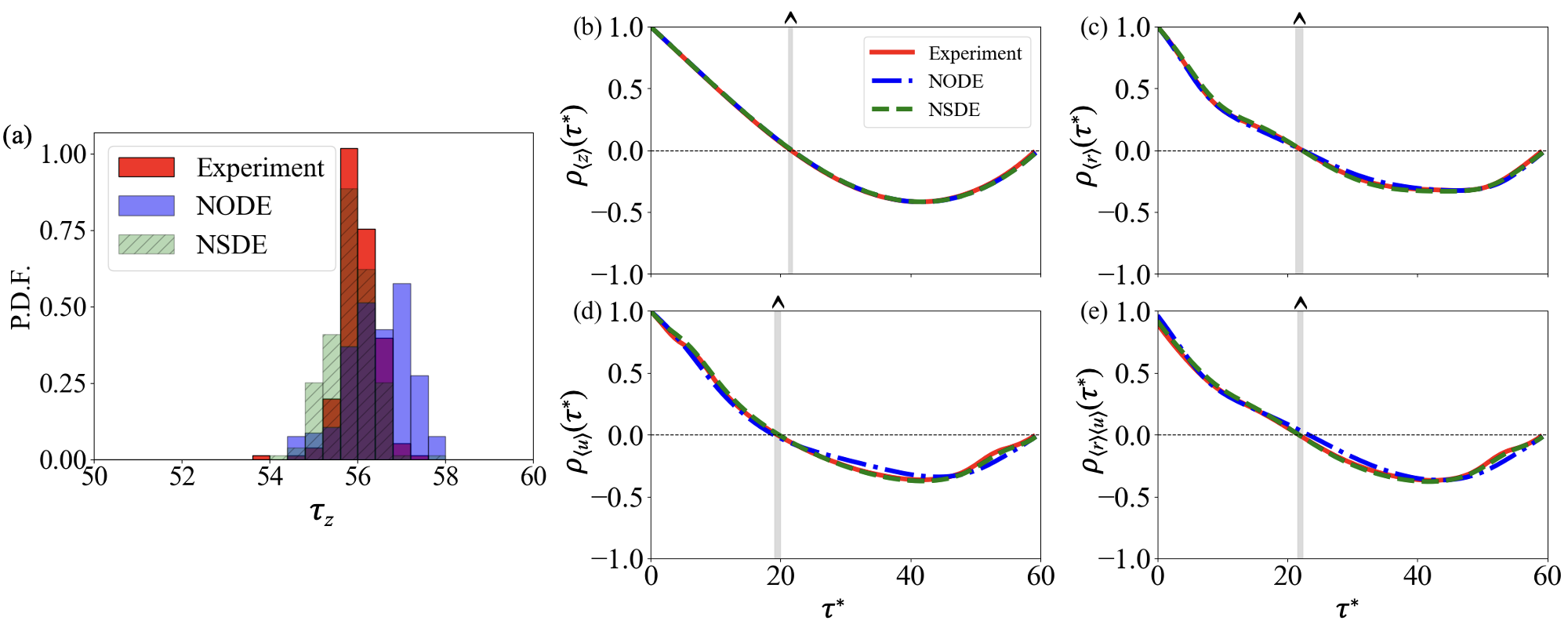}
    \caption{(a) Probability density distribution of the dimensionless settling time $\tau_z$, with mean values of 55.8, 56.2, and 55.6 for the experiment, NODE, and NSDE predictions. (b-d) Auto-correlation coefficients of ensemble-averaged depth $\langle z \rangle$, lateral displacement $\langle r \rangle$, and streamwise velocity $\langle u \rangle$ as a function of the non-dimensional time lag $\tau^*$. (e) Cross-correlation coefficients between $\langle r \rangle$ and $\langle u \rangle$.}
    \label{settling_corr}
\end{figure}

\begin{figure}
    \centering
    \includegraphics[width=1\linewidth]{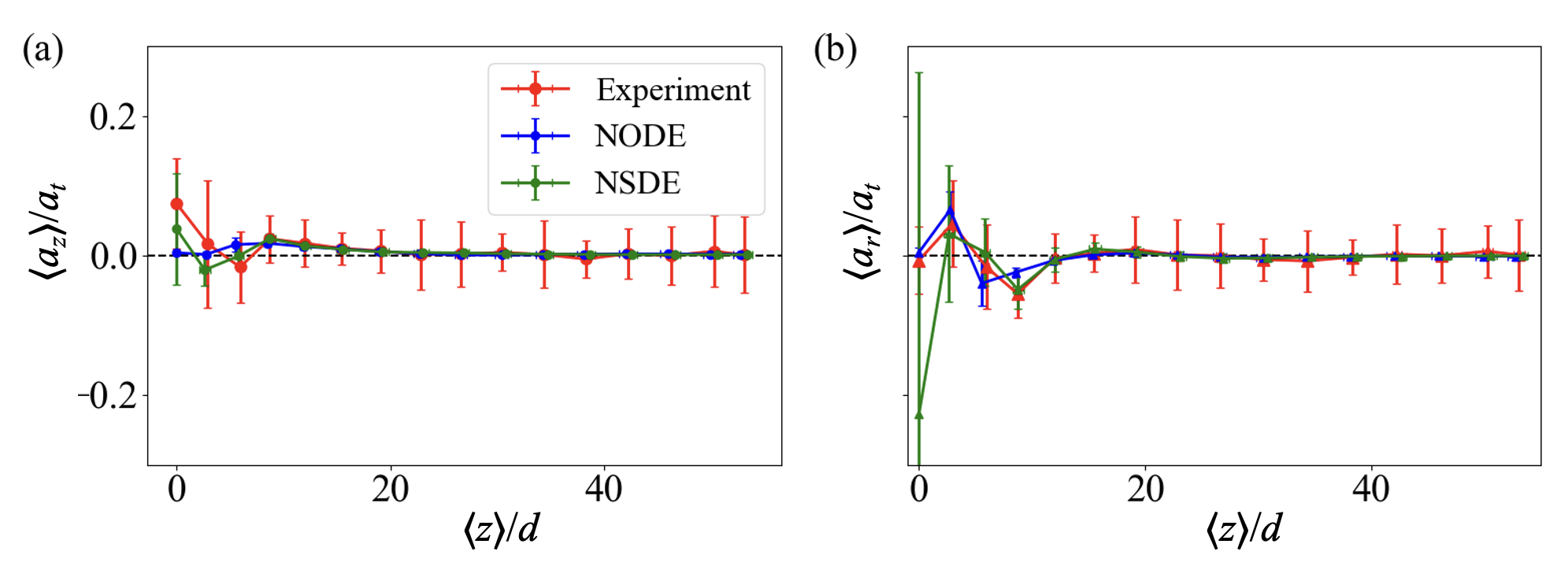}
    \caption{Ensemble-averaged acceleration components for the NODE and NSDE models, and experimental data for the streamwise component $\langle a_z \rangle$ and lateral component $\langle a_r \rangle$, corresponding to panels (a) and (b), respectively.
    Error bars indicate standard deviations. \label{acc}}
    
\end{figure}

For the non-dimensional representation of velocity components, we consider both theoretical and experimental estimates of the terminal velocity. The theoretical value follows the expression proposed by \citet{goossens2020new}:  
\begin{equation}  
U_t=\frac{[(729+3A_r)-27]^{\frac{1}{2}}\mu}{\rho_f d} \approx 363 \text{ mm/s},  
\end{equation}  
where $A_r=\rho_f(\rho_s - \rho_f)gd^3/\mu^2=2.04\times10^6$ is the Archimedes number. This prediction closely matches the experimentally measured value of $U_t \sim 360$ mm/s. Given this agreement, we adopt the theoretical value as the reference for the ensemble-averaged terminal velocity.

Fig. \ref{means}a presents the dimensionless ensemble-averaged streamwise velocity $\langle u \rangle / U_t$ and lateral velocity $\langle v \rangle / U_t$ from the experimental data and NODE/NSDE predictions. The results indicate that the sphere attains its terminal velocity at approximately $\langle z \rangle / d \approx 40$, where the lateral velocity also approaches zero. The NODE model closely captures the trends in velocity evolution for both the streamwise and lateral components across different depths, ultimately converging to the right terminal velocity. However, some deviations between the NODE reconstruction and the experimental data are observed in the initial stages (e.g., see $\langle u\rangle/U_t$ when $\langle z\rangle/d<6$ in Fig. \ref{means}a). These discrepancies likely arise from two reasons: the nonlinear transient dynamics inherent to the system, which are not fully accounted for in the model's simplifications, and the fact that the model trained on more data from the later stages, may not fully capture the dynamics during the earlier stages of the trajectories. We note additionally that there are large gradients in the NSDE reconstructed lateral velocities (e.g., see $\langle v\rangle/U_t$ when $\langle z\rangle/d<4$ in Fig. \ref{means}a). This is caused by a large jump and drop in the lateral profile past the initial condition, which leads to steep positive and negative velocities. To further address this point, one can see that a small set of reconstructed trajectories in Fig.~\ref{lateral_pdf}c exhibits significant lateral displacements immediately after \(t^*=0\) which are much larger than those observed in Fig.~\ref{lateral_pdf}a, indicating an unrealistic overshoot near the start of the trajectories. As a result of this, some trajectories that remain closer to the initial transient region display very steep gradients, resulting in large velocity magnitudes in this area. Despite these fluctuations, the large shift between positive and negative magnitudes suggest that, although these portions are unrealistic, that the profiles generated by the NSDE model tend to converge over time toward the mean velocity profile of the trajectories, both in the later transient phase around $t^*\approx 15$ and in the settling phase as $t^*\rightarrow60$. This is shown further by the fact the \(\langle r\rangle/d\) in Fig. \ref{means}b has the second reconstructed point as the only significant deviation, with a noticeably sharper gradient than the ground truth data. A key contributor to this is shown in Fig. \ref{lateral_pdf}c where a set of trajectories jump to $r/d\sim 1.5$ within just a few timesteps.
The reconstructed $\langle u\rangle/U_t,$ exhibits small error bars, as can be seen in Fig. \ref{means}a. One can see that the experimental data exhibits the same statistics due to the determinism inherent in the streamwise velocity which is primarily affected by gravitational forces. Of note is the fact that when reconstructing the full dataset the model accurately predicts the statistics  of the trajectories at nearly all points past the transient region, as shown by the error bars in Fig. \ref{means}b. As a result, we see that the models capture both low and high variance modes effectively.

We also observe that the reconstructed $\langle v\rangle/U_t$ near the settling time appear significantly smoother than those of the ground truth data ( Fig. \ref{means}a). However, the NSDE predictions retain noticeably more of the stochastic nature of the true dynamics  compared to those produced by the NODE. This behavior is  typical in temporal prediction using NODEs, where high frequency variations in the  data are normally smoothed during training. This smoothing effect arises from the difficulty of capturing fine-scale details during backpropagation. One potential solution is to employ a larger network; however, this comes at the cost of significantly increased computational requirements. Therefore, a remarkable advantage of the NSDEs are their ability to recover fine scale  details  despite using a  small network. NSDEs  also  faithfully reproduce  the statistical behavior of the system.

We next investigated the impact of dataset size on the accuracy of the reconstructed trajectories, focusing on \(\langle r\rangle\) and  \(\langle z\rangle\) from model predictions. Thus, we trained both NODE and NSDE models with identical architectures using three different dataset sizes: 100$\%$ ($\approx$200 trajectories), 50$\%$ (100 trajectories), and 10$\%$ (20 trajectories). The models trained on the full dataset provides the most accurate reconstruction, closely matching the experimental ground truth (see Fig. \ref{means}b).  This is because the larger dataset offers a more comprehensive representation of the dynamics of the system, allowing the models to capture both global trends and finer details. The models trained on 50$\%$ of the data captures the general trend of the ensemble-averages; however, significant deviations appear at both the initial and terminal stages (see Fig. \ref{means}c when \(\langle z\rangle/d <18\) and \(\langle z\rangle/d > 20\)). This is likely because the reduced dataset does not provide sufficient information to accurately represent the system's dynamics. When trained on only 10$\%$ of the data (see Fig. \ref{means}d), the models fail in different ways: NODE fails to learn the underlying trend accurately and instead converges to underrepresented trajectories, whereas NSDE predicts a trend with significant offsets from the experimental trend, as the limited dataset lacks the diversity necessary for models to generalize properly.
Given these observations, we select the models trained on the full dataset for further analysis (if not stated otherwise) to ensure the most reliable and accurate results, as they captures the complete regime of the system dynamics.

To better understand the capabilities of both the NODE and NSDE models,  we also evaluate their predictions on the settling time, comparing them with  experimental results. The settling time is defined as the duration between the particle’s release moment and when its depth reaches 50 times of its diameter:
\begin{equation}
\tau_z = \frac{U_t}{d} t \Big|_{z/d=50}.
\end{equation}
Both the NODE and NSDE models demonstrate excellent agreement with experimental data in the settling region, with mean settling time of the reconstructed trajectories measured as $\bar\tau_z =56.2$ and $\bar\tau_z =55.6$, respectively. These closely match the experimental mean settling time ($\bar\tau_z= 55.8$), resulting in relative errors of only $0.7\%$ and $0.4\%$. The corresponding probability distributions from both models  and  experimental data are shown in Fig. \ref{settling_corr}a.

Finally, Fig. \ref{settling_corr}b-e show the auto- and cross-correlation coefficients among several  ensemble-averaged properties: \(\langle z\rangle\), \(\langle r\rangle\) and \(\langle u\rangle\). The auto- and crosscorrelation coefficients are defined respectively as:
\begin{equation}
    \rho_{\langle p\rangle}(\tau^*)=\frac{1}{\sigma_{\langle p\rangle}^2 }\int_{0}^{T^*} \langle p\rangle(t^*)\langle p\rangle(t^*+\tau^*)dt^*
\end{equation}
\begin{equation}
    \rho_{\langle p\rangle \langle q\rangle}(\tau^*)=\frac{1}{\sigma_{\langle p\rangle} \sigma_{\langle q\rangle}}\int_{0}^{T^*} \langle p\rangle(t^*)\langle q\rangle(t^*+\tau^*)dt^*
\end{equation}
where $p$ and $q$ refer to the properties of interest, $\tau^* = \tau U_t / d$ is the dimensionless time lag, $T^*$ is the total time lag range, $\sigma_{\langle p \rangle}$ and $\sigma_{\langle q \rangle}$ are the standard deviations of $\langle p \rangle$ and $\langle q \rangle$. These correlations provide complementary insights into the temporal relationship between the settling process and flow dynamics. We note that the correlation functions generated by both the NODE and NSDE reconstruction agree well with those of the experimental data. The autocorrelations $\rho_{\langle z\rangle}(\tau^*)$, $\rho_{\langle r\rangle}(\tau^*),$ and $\rho_{\langle u\rangle}(\tau^*)$ of NODE and NSDE predictions decay to zero in nearly the same ranges as those of the experimental data (see Fig. \ref{settling_corr}b-d). Beyond this point indicated by the gray bars, the particle eventually reaches a steady terminal velocity with negligible lateral motion. Additionally, the cross-correlation function $\rho_{\langle r \rangle \langle u \rangle}(\tau^*)$ decays to zero within the range of $\tau^* \in [21.5, 22.4]$ (indicated by the gray bar in Fig. \ref{settling_corr}e), marking the end of the transient phase and the onset of the long-term settling process.

To further investigate the robustness of both models in accurately capturing the system dynamics, we compute the acceleration from the  trajectories. Since acceleration represents second-order dynamics, we apply a numerical scheme to derive it from the time-resolved position data. 
We used a second-order central difference scheme to compute the accelerations from the particle trajectories as
\begin{equation}
   \begin{bmatrix}
a_z \\
a_r \\
\end{bmatrix} =\mathbf{a}(t) \approx \frac{\mathbf{x}(t +\delta t)-2\mathbf{x}(t)+\mathbf{x}(t-\delta t)}{\delta t^2},\mathbf{x}=\begin{bmatrix}
z \\
r \\
\end{bmatrix}
\end{equation}
where $a_z$ and $a_r$ are the accelerations in the streamwise ($z$) and lateral ($r$) directions. Figure \ref{acc} presents the ensemble-averaged streamwise acceleration $\langle a_z \rangle$ and lateral acceleration $\langle a_r \rangle$ as functions of $\langle z \rangle/d$, both normalized by the characteristic acceleration $a_t = U_t^2/d$. The experimental data exhibit oscillations in both accelerations at early times (or positions) as the particle penetrates the air-liquid interface, eventually approaching zero at later stages as it reaches a steady state.  As shown in Fig. \ref{acc}a, the NODE model accurately reproduces the ensemble-averaged accelerations, though discrepancies arise in the variances, where the model tends to underestimate fluctuations. A similar trend is observed for the NSDE model in Fig. \ref{acc}b, where short-time initial transients are not well captured. However, the long-time dynamics are reconstructed with minor error, except for an underestimation of variance. 

Note that the error bars of the predicted accelerations in Fig. \ref{acc}a-b are small, particularly past the transient region. This may, in part, be attributed to the lower variability inherent in the NSDE predictions, particularly near the settling region, where the acceleration error bars are both large and highly time-dependent. 
The trendline for the lateral acceleration component does not fully capture the subtle oscillatory behavior observed at early times. 
This may also be attributed to the fact that the particles tend to oscillate in a manner that is not fully captured by the NSDEs across many trajectories. In particular, a close inspection of Fig. \ref{lateral_pdf}a-c reveals  that the NSDE model exhibits a slight preference for learning trajectories whose lateral displacements deviate less. The same result holds for the NODEs, however, the error bars in \ref{acc}a-b tend to be marginally larger, and the oscillation in the mean value of acceleration is captured more accurately.

\section{Conclusions}
We developed data-driven models to predict the dynamics of a spherical particle settling in a quiescent Newtonian medium. Instead of explicitly modeling the underlying fluid dynamics, we learn the particle’s spatial evolution. Using PTV, we tracked the particle’s motion from its initial release right above the free surface to past the terminal velocity, providing detailed, time-resolved trajectory data. With this experimental dataset, we trained neural networks with both deterministic (NODE) and stochastic (NSDE) approaches. We demonstrated that NODE excels at generalizing across unlimited unseen data, accurately  reconstructing the collective behavior  as long as the unseen initial conditions maintain the same distribution to those of the training data. The NODE model can accurately predict the terminal velocity and settling time, with excellent agreement to the experimental data. Also, we demonstrate that NODE can provide statistically accurate predictions, with some errors, even when trained on only half of the dataset. The comparison of the results between the two models opens the question of the determinism of the flow regime captured by the experimental data. While the NODE model produces smooth trajectories,  the experimental data exhibit irregular lateral velocity fluctuations. This contrast suggests that, at the length scales and Reynolds number considered, the flow may possess an intrinsic stochastic component, rather than being fully described by deterministic dynamics alone.

On the other hand, NSDE can quickly outline the collective behavior even being trained with a much smaller dataset, but cannot generalize the short-time dynamics of unseen initial conditions given the exclusive dependence between each trajectory and its own cubic spline coefficients. Its generalization lies specifically in its ability to map a set of these cubic spline coefficients to a statistically similar set of predictions. Nevertheless, both models can accurately forecast the statistical characteristics of the settling sphere system up to the second-order dynamics with minor error, despite the system’s inherent complex dynamics. 

Now that we have demonstrated that our approaches can model the settling of particle in quiescent Newtonian flow,  several avenues  remain open for future research. One of them is the generalization of these models to a wider range of fluid fields (such as shear flow and extensional flow), fluid properties (including non-Newtonian fluids and those with temperature or concentration gradients), and object shape dependence. One particularly promising avenue is exploring the dispersion and interaction between multiple particles. The dynamics of multiple interacting particles can introduce a higher level of complexity, including collective behavior, collision dynamics, and potential changes in settling velocity. By data-driven modeling these interactions, we can gain deeper insights into particle behavior in suspensions or granular flows, which are relevant for many industrial and environmental applications, such as sediment transport, filtration, and material processing.



\appendix

\section*{Appendices}

\subsection*{A1. NODE Prediction on test data \label{test_data_app}}

\begin{figure}
    \centering
    \includegraphics[width=1\linewidth]{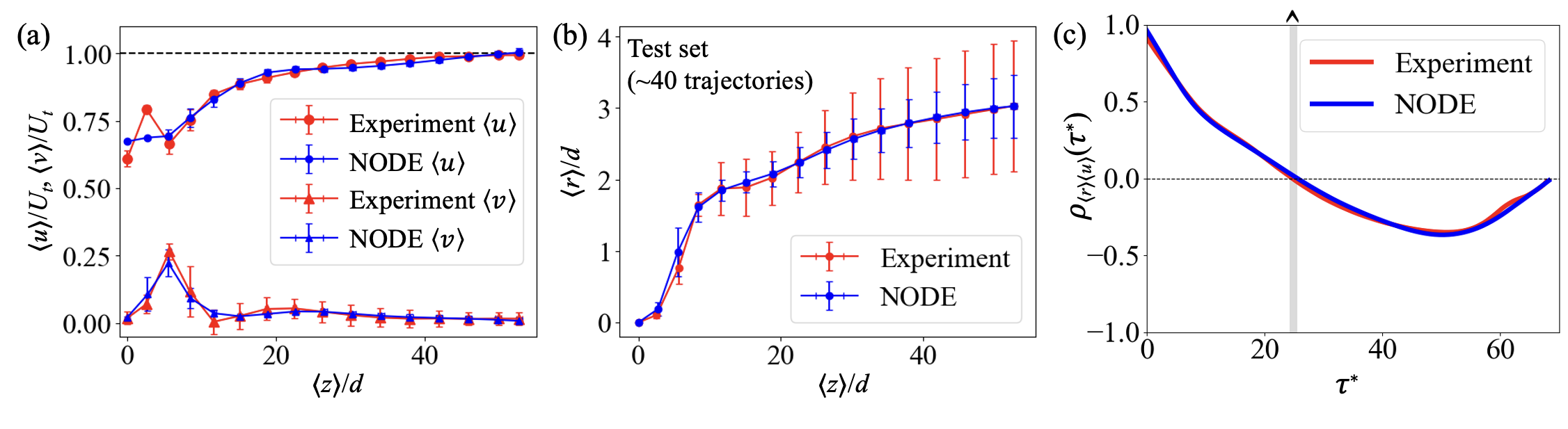}
    \caption{  (a) Ensemble-averaged streamwise and lateral velocities, (b) ensemble-averaged lateral displacement, and (c) crosscorrelation between $\langle r \rangle$ and $\langle u \rangle$ for the test set. The results demonstrate that the proposed model accurately captures the collective behavior of the test dataset, highlighting its robustness and predictive capability.}
    \label{test}
\end{figure}

Here, we evaluate the NODE models using initial conditions from the experimental test dataset.  The 189 ground truth trajectories obtained from PTV  are  are randomly split into $80\%$ for training and $20\%$ for testing.
Fig. \ref{test} shows the ensemble-averaged velocities, lateral displacement, and cross-correlation coefficients on the test dataset, compared to the experimental data. It shows that the data-driven model is capable of successfully capturing the velocity profiles and lateral displacement, with an excellent match in the cross-correlations, even for initial conditions that were not seen during training. 



\bibliography{biblio}

\end{document}